\documentstyle[12pt]{article}

\title{On quantum properties of the four-dimensional generic chiral superfield model}

\author{A.T. Banin$^1$\footnote{atb@math.nsc.ru}, I.L.
Buchbinder$^2$\footnote{joseph@tspu.edu.ru}, N.G.
Pletnev$^1$\footnote{pletnev@math.nsc.ru}}

\date{\it
$^1$Department of Theoretical Physics\\ Institute of Mathematics,
Novosibirsk, \\ 630090, Russia\\$^2$ Department of Theoretical
Physics\\ Tomsk State Pedagogical University\\ Tomsk 634041,
Russia }

\begin{document}
\maketitle

\begin{abstract}
We study a problem of systematical evaluation of the quantum
corrections for general $4D$ supersymmetric K\"ahler sigma models
with chiral and antichiral superpotentials. Using manifestly
reparametrization covariant techniques (the background-quantum
splitting and proper-time representation) in the ${\cal N}=1$
superspace we show how to define unambiguously the one-loop
effective action. We introduce the reparametrization covariant
derivatives acting on superfields and prove that their algebra is
analogous to algebra in super Yang-Mills (SYM) theory. This
analogy allows us to use for evaluation of the effective action in
the theory under consideration methods developed for SYM theory.
The divergencies for the model are obtained. It is shown that on
general K\"ahler manifold the one-loop counterterms have the
structure of supersymmetric WZNW term in the form proposed in Ref.
\cite{swzw}. Leading finite contribution in covariant derivative
expansion of the one-loop effective action (superfield $a_3$
coefficient) is calculated.

\end{abstract}

\thispagestyle{empty}
\newcommand{\be}{\begin{equation}}
\newcommand{\ee}{\end{equation}}
\newcommand{\bea}{\begin{eqnarray}}
\newcommand{\eea}{\end{eqnarray}}

\section{Introduction}

Nonlinear sigma models play a significant role in many areas of
the field theory. An important class of such models is presented
by $2D$ conformal field theories, which are exactly solvable (in a
sense that $S$-matrix, the correlation functions of the different
fields and the anomalous dimensions are completely determined on
the base of conformal invariance) \cite{bel} and consistent at the
quantum level \cite{1}. Two-dimensional supersymmetric nonlinear
sigma-models possess many remarkable properties. They are
renormalized field theories. Moreover for a field theory on the
world sheet the correlations between conformal invariance,
(extended) supersymmetry and geometry of the complex manifolds of
the full quantum theory yield to powerful restrictions on the
background fields geometry in each order of the perturbation
theory \cite{5}.

The remarkable properties of supersymmetric $2D$ nonlinear
$\sigma$-models inspired interest to constructing the geometrical
non-polynomial theories of the supersymmetric matter in $4D$
space-time and studying their properties \cite{7}, \cite{ket}.
Supersymmetric nonlinear sigma-models have been formulated both
for simple and for extended supersymmetries (see e.g. \cite{bk},
\cite{14}, \cite{N2} for review). It was proved that in $4D$ the
target space of rigid supersymmetric nonlinear $\sigma$-models
must be the K\"ahler manifold \cite{z} for ${\cal N}=1$
supersymmetry and hyper-K\"ahler manifold \cite{1}, \cite{5} for
${\cal N}=2$ supersymmetry. A number of supersymmetric
sigma-models has been constructed within superstring theory
\cite{GSW} where the extra dimensions are wrapped up into a coset
space. Unlike $2D$ models, the $4D$ nonlinear supersymmetric
sigma-models are nonrenormalizable in power-counting as well as
their $4D$ nonsupersymmetric predecessors. This is a main reason
why the quantum aspects of such models are not well studied (see
however some attempts in Refs. \cite{bcp}).

Another large class of nonlinear sigma-models is formed by the
$4D$ low-energy effective phenomenological theories. Dynamics of
these models is invariant under a global group ${\cal G}$, whereas
a vacuum is invariant only under some subgroup ${\cal H}$
\cite{wein}. It is well known that such models are
nonrenormalizable under a power-counting analysis and requires the
introduction of new couplings in each order of the loop expansion.
However, the higher order loop terms involve higher powers of the
momentum, and thus, the low energy behavior is controlled only by
the lower order terms which are unambiguous and do not undergo
further renormalization (see e.g. \cite{Mary}). These models
possess the global anomalies which can be reproduced at the
low-energy scale by the four-derivative Wess-Zumino-Novikov-Witten
(WZNW) action \cite{an}.

Manifest $N=1$, $4D$ superfield form of the WZNW term has been
considered in \cite{nr} but the proposed construction requires an
infinite number of unspecified constants that appear in an
undetermined function $\beta_{ij\bar{k}}$. In addition the
auxiliary fields become propagating fields. The alternative form
of the $4D$, ${\cal N}=1$ ungauged supersymmetric WZNW model has
been given in \cite{swzw}. The interesting superfield sigma-models
using the CNM (chiral/non-minimal) formulation have been
constructed in \cite{3} where chiral superfields exist in a tandem
with complex linear superfields. The physical superfields $\Phi,
\bar{\Phi}$ and $\Gamma, \bar\Gamma$ in these models should be
regarded, respectively, as a coordinates of the K\"ahler manifold
and a tangent vector at a point $\Phi, \bar\Phi$ of the same
manifold \cite{kuz}. The models are closely related to
phenomenological models of pion physics and admit interpretations
as low energy string-like models associated with QCD.  Supergraph
technique for these models was given in Ref. \cite{cnm}. Next
natural step here should be a development of a background field
expansion in terms of normal coordinates and constructing of a
computational procedure for finding counterterms and finite
contributions to the effective action (see the references for
early literature in \cite{6} and \cite{how}). This is one of our
motivations to study quantum properties of the K\"ahler sigma
model, which is a part of CNM sigma models.

One-loop divergencies for the standard ${\cal N}=1$ and ${\cal
N}=2$ supersymmetric nonlinear sigma models on the K\"ahler and
hyper-K\"ahler manifolds respectively demonstrate an existence of
divergent terms with four order derivatives of background fields.
The origin of such divergences is the nonrenormalizability of the
model, and, therefore, to make quantum theory multiplicatively
renormalizable we have to write terms with four order derivatives
in the action from the very beginning. Such a situation is
analogous e.g. to the relation between nonrenormalized Einstein
quantum gravity and asymptotically free renormalized $R^2$ quantum
gravity (see e.g.\cite{8}). Unfortunately a detailed analysis of
quantum properties of generic $4D$ ${\cal N}=1, 2$
supersymmetrical models with some set of chiral supermultiplets
has not been carried out so far.

The goal of this paper is to describe quantum aspects of the the
$4D$ generic chiral superfield model including supersymmetric
sigma-model as a particular case. To be more precise, we formulate
the heat kernel approach for the covariant computations of the
one-loop effective action in the $4D$ generic chiral superfield
model. Modern interest to this problem was inspired by recent
development of generic chiral superfield models on
nonanticommutative (NAC) superspaces (see e.g. \cite{9D2},
\cite{9}, \cite{9r}). Classical structure of such models has been
thoroughly studied while their quantum properties requires a
further analysis. In addition to the problems inherent with
nonanticommutative models there is a known general difficulty in
superfield sigma-models: the absence of chiral and simultaneously
holomorphic normal coordinates on the generic K\"ahler manifolds
does not allow to develop a loop expansion preserving all
symmetries of the theory. This fact has been already mentioned in
the pioneering papers \cite{1}, \cite{5}. Some papers were
directly addressed to treatment of the above difficulty \cite{11},
\cite{11n}. Unfortunately this difficulty can not be overcome in
general since neither chiral metric nor chiral Levi-Civita
connection do not exist on the K\"ahler manifold (even having
isometries). Therefore the geodesics also do not exist in a
subspace of chiral superfields and we can not utilize an expansion
on a non-trivial superfield background in a way which preserves
the chirality of the model. However as it has been pointed out
some time ago (see e.g.\cite{ket}) this problem is unessential for
one-loop calculations because the deviation from chirality
$\bar{D}\sigma^i \sim {\cal O}(\sigma^2)$ is quadratic over
quantum fluctuations and therefore the above difficulty arises
only in the higher loops.

The paper is organized as follows. In section 2 we present some
mathematical grounds related to the model under consideration and
discuss classical properties of the model. In section 3 we
consider the normal coordinate expansion for the K\"ahler
potential and superpotentials. Then we introduce the specific
covariant derivatives, formulate an algebra of these derivatives,
study their basic properties and observe the analogies with SYM
theories. As a result, we prove that the $4D$ superfield
sigma-models are characterized by the objects which are analogues
to superfield strengths in SYM theory. These objects have
well-definite transformation properties and naturally arise as the
building blocks for constructing the effective action. In section
4 we fulfil the one-loop calculations and find the divergencies
and some finite corrections. Discussion and conclusions are given
in section 5.

\section{Generic chiral superfield model}

In this section we briefly discuss the basic notions of the generic
chiral multiplet model in superspace (see e.g. \cite{bk}) which will
be used further.

The model under consideration is a map from ${\cal N}=1$ superspace
into the K\"ahler space and is described by chiral $\Phi^i$ and
antichiral $\bar\Phi^{\bar{i}}$ superfields whose components, the
complex scalars $\phi^i, \bar\phi^{\bar{i}}$, play the roles of
coordinates on the K\"ahler manifold, whereas the fermions $\psi^i,
\bar\psi^{\bar{i}}$ transform as vectors on a target manifold. The
corresponding superspace action is written in terms of the K\"ahler
potential $K(\Phi, \bar{\Phi})$ and chiral and antichiral
superpotentials $P(\Phi), \bar{P}(\bar\Phi)$ and has the form

\begin{equation}\label{kahl_act}
S=\int d^8z \,K (\Phi, \bar{\Phi}) +\int d^6z \,P(\Phi) +\int d^6
\bar{z} \,\bar{P}(\bar\Phi)~.
\end{equation}
This action is invariant under holomorphic reparametrization of
superfields
\begin{equation}\label{repar}
\Phi^i \rightarrow
\Phi^{i\; '}=f^i(\Phi), \quad \bar\Phi^{\bar{i}} \rightarrow
\bar\Phi^{\bar{i}\; '}=\bar{f}^{\bar{i}}(\bar\Phi)~.
\end{equation}
and the K\"ahler transformations
\begin{equation}\label{gaug}
K(\Phi,\bar\Phi)\rightarrow K(\Phi,\bar\Phi)
+F(\Phi)+\bar{F}(\bar\Phi)~,
\end{equation}
with some $F, \bar{F}$. The K\"ahler potential $K(\Phi,\bar{\Phi})$
defines the metric of the target manifold
\begin{equation}
g_{i\bar{j}}=\frac{\partial^2 K(\Phi,\bar{\Phi})}{\partial \Phi^i
\partial \bar{\Phi}^{\bar{j}}} \equiv K_{,\, i\bar{j}}(\Phi,
\bar{\Phi})~.
\end{equation}
This metric allows to find the components of the Levi-Civita
connection
\begin{equation}
\Gamma^i_{kl}=g^{i\bar{m}}
\partial_l g_{\bar{m}k}, \quad
\Gamma^{\bar{i}}_{\bar{k}\bar{l}}=g^{\bar{i}m}
\partial_{\bar{l}} g_{m\bar{k}}, \quad g_{i\bar{j}\, ;\, k}=g_{i\bar{j}\,;\,
\bar{k}}=0~.
\end{equation}
Using the above connection we introduce the standard covariant
derivatives of the geometrical objects with target space indices
\begin{equation}\label{kcov}
A^i_{; l}=\partial_l A^i +\Gamma^i_{kl}A^k ,\quad A^{\bar{i}}_{;
l}=\partial_l A^{\bar{i}}~, \quad A^{i}_{;
\bar{l}}=\partial_{\bar{l}} A^{i},\quad A^{\bar{i}}_{;
\bar{l}}=\partial_{\bar{l}} A^{\bar{i}}
+\Gamma^{\bar{i}}_{\bar{k}\bar{l}}A^{\bar{k}}~.
\end{equation}

The Lagrangian for the component fields is obtained as the lowest
component of the superfield ${\cal L}=D^\alpha \bar{D}^2D_\alpha K$.
The component form of the K\"ahler $D$-term is written from the
expression
\begin{equation}\label{class}
K|_{D}= -i g_{i\,\bar{j}}\partial^{\alpha\dot\alpha}\Phi^i
\partial_{\alpha\dot\alpha}\bar{\Phi}^{\bar{j}}-g_{i\,\bar{j}}({\cal D}_{\alpha\dot\alpha}D^\alpha
\Phi^i)\bar{D}^{\dot\alpha}\bar{\Phi}^{\bar{j}}+g_{i\,\bar{j}}D_\alpha
\Phi^i ({\cal
D}^{\alpha\dot\alpha}\bar{D}_{\dot\alpha}\bar{\Phi}^{\bar{j}})
\end{equation}
$$
+2i g_{i\,\bar{j}}\bar{\cal F}^{\bar{j}}{\cal F}^i
+\frac{i}{2}R_{i\,\bar{l}\, \bar{k} \,m}D^\alpha \Phi^mD_\alpha
\Phi^i
D^{\dot\alpha}\bar{\Phi}^{\bar{k}}D_{\dot\alpha}\bar{\Phi}^{\bar{l}}~.
$$
Here we have introduced the reparametrization covariant derivatives
\begin{equation}\label{cccov}
{\cal D}^\alpha A^i =D^\alpha A^i +\Gamma^i_{l\,k}D^\alpha \Phi^k
A^l,\quad {\cal D}^{\dot\alpha}\bar{A}^{\bar{j}}=
{D}^{\dot\alpha}\bar{A}^{\bar{j}}+\Gamma^{\bar{j}}_{\bar{l}\,\bar{k}}D^{\dot\alpha}
\bar{\Phi}^{\bar{k}}\bar{A}^{\bar{l}}~,
\end{equation}
which will be important ingredients for constructing the effective
action. The curvature tensor of the K\"ahler manifold has the form
$R_{\bar{i}kl\bar{m}}=-g_{\bar{i}\,j}\partial_{\bar{m}}\Gamma^j_{kl}$
and we use following definitions for combinations of the auxiliary
and spinoral fields  that transform as tangent vectors
\begin{equation}{\cal F}^i =\frac{1}{2}{\cal D}^\alpha D_\alpha
\Phi^i, \quad \bar{\cal F}^{\bar{i}}=\frac{1}{2}{\cal
D}^{\dot\alpha}D_{\dot\alpha} \bar{\Phi}^{\bar{i}}~.
\end{equation}
Projection of (\ref{class}) to the lowest superfield components
results  the most general $4D$ ${\cal N}=1$ second-order Lagrangian
\cite{bk}, \cite{14} in the component form. The F-term
superpotential defines of the scalar potential and Yukawa couplings.

The equations of motion for the model (\ref{kahl_act}) in superfield
form are given by
\begin{equation}\label{eq_m}
\bar{\cal D}^2 K_{,\,i}=P_{, \;i} ,\quad {\cal D}^2
K_{,\,\bar{i}}=\bar{P}_{,\;\bar{i}}~.
\end{equation}

In section 4 we study the one-loop effective action for the model
(\ref{kahl_act}).

\section{Background-quantum splitting}

Covariant calculation of the quantum corrections for the model
(\ref{kahl_act}) is based on the loop expansion with help of
background-quantum splitting which preserves the symmetry of the
model. For construction of such a splitting it is useful to
compare the superfield $\sigma$-model with conventional
$\sigma$-model. The background-quantum splitting for conventional
$\sigma$-model is realized in terms of Riemann normal coordinates
what allows to retain the general coordinate invariance(for
references in early literature, see e.g. \cite{1}, \cite{6}).
Therefore the Riemann normal coordinates are the basic ingredient
in constructing covariant loop expansion for $\sigma$-model
effective action due to the following property: the geodesics
passing through the origin have the same form $d^2
y^i/d\lambda^2=0$ (here $\lambda$ is a affine parameter, "time"
along the geodesic) as the equations of straight lines passing
through the origin of a Cartesian system of coordinates in a flat
geometry.

The superfield $\sigma$-models are associated with K\"ahler
geometry which possess a complex structure. The Riemann normal
coordinates mix the holomorphic and antiholomorphic coordinates
and hence violate the reparametrization symmetry since the set of
holomorphic coordinate transformations (\ref{repar}) is only
subset of a full set of general coordinate transformations on
manifold parametrized by the coordinates ${\Phi}^{i}$ and
$\bar{\Phi}^{\bar{i}}$. As a result, a general covariant loop
expansion of the effective action preserving the complex structure
is impossible in principle.

The problem of building a manifestly covariant background-quantum
splitting for supersymmetrical $\sigma$-model was discussed by a
number of authors (see e.g. \cite{11}, \cite{11n}). In this
section we compare two various approaches to this problem and
demonstrate that for one-loop calculations such a problem does not
exist really.

\subsection{Chiral coordinate expansion}
First of all we shortly discuss a well-known decomposition (see e.g.
\cite{11}, \cite{11n}) of the complex (anti)chiral superfields into
background superfields $\phi^i, (\bar\phi^{\bar{i}} )$ and quantum
superfields $\pi^i, (\bar\pi^{\bar{i}})$ fluctuating around them:
\begin{equation}\label{lamb_exp}
\Phi^i(z)=\phi^i(z)+\pi^i(z), \quad
\bar\Phi^{\bar{i}}(z)=\bar\phi^{\bar{i}}(z)+\bar\pi^{\bar{i}}(z)~.
\end{equation}
In (\ref{lamb_exp}) the quantum fields $\pi^i (\bar\pi^{\bar{i}})$
are differences between coordinates on the manifold and therefore
do not transform as the vectors under general coordinate
transformations. Hence the expansion of the geometric objects on
the manifold in a power series in $\pi (\bar\pi)$ will not be
covariant at each order. Instead we consider the velocity vectors
$\sigma^i(\pi), \bar\sigma^{\bar{i}}(\bar\pi)$ which are chosen to
play the role of the quantum fields. According to Ref. \cite{11}
we consider an affine parameter $\lambda$ ($0\leq \lambda \leq 1$)
along an arbitrary path from $\Phi(0)=\phi$ to $\Phi(1)=\phi +\pi$
and expand the field in the chiral coordinates
\begin{equation}\label{bphiexp}
\Phi^i(\lambda)=\phi^i
+\sum^\infty_{n=1}\frac{\lambda^n}{n!}\phi^i_{(n)}~.
\end{equation}
Coordinate transformations on the curve are described by
$\delta\Phi^i(\lambda)=f^i(\Phi(\lambda))$. Because of the
transformation rule $\delta \frac{\partial \Phi^i(\lambda)}{\partial
\lambda}=f^i_{,\; j}(\Phi(\lambda))\frac{\partial
\Phi^i(\lambda)}{\partial \lambda}$ the quantity like
$\frac{\partial \Phi^i(\lambda)}{\partial \lambda}$ is a
contravariant chiral tangent vector at each value of $\lambda$,
including the quantity $\sigma^i=\frac{\partial
\Phi^i(\lambda)}{\partial \lambda}|_{\lambda=0}$. The coefficients
$\phi^i_{(n)}, (\bar\phi^{\bar{i}}_{(n)})$ can be obtained directly
from analysis of the geodesics equation
\begin{equation}\label{geod}
\frac{d^2\Phi^i}{d\lambda^2} +\Gamma^i_{jk}(\Phi,
\bar\Phi)\frac{d\Phi^j}{d\lambda}\frac{d\Phi^k}{d\lambda}=0~,
\end{equation}
which, of course, is incompatible with the chirality condition on
$\Phi^i$, since the $\Gamma^i_{jk}$ depends on both $\Phi^i$ and
$\bar\Phi^{\bar{i}}$. Substituting the expansion in affine parameter
for the curves (\ref{lamb_exp}) to above equation, we find the
recursive formulae for the coefficients $\phi^i_{(n)},
(\bar\phi^{\bar{i}}_{(n)})$. For example
\begin{equation}
\phi^i_{(2)}=-\Gamma^i_{jk}\sigma^j \sigma^k, \quad
\phi^i_{(3)}=-(\Gamma^i_{jk,l}
-2\Gamma^i_{jm}\Gamma^m_{kl})\sigma^j\sigma^k\sigma^l~.
\end{equation}

In the this paper we concentrate only on one-loop analysis where
we need only in the second order in an expansion of the classical
action in quantum superfields. It is well-known (see e.g.
\cite{11}) that Taylor expansion of the K\"ahler potential in
normal coordinates has the form
\begin{equation}\label{kex}
\begin{array}{l}
K(\phi+\pi,\bar\phi+\bar\pi)=K(\phi,\bar\phi)+K_{,\,i}\sigma^i
+K_{,\,\bar{i}}\bar\sigma^{\bar{i}} \\
+\frac{1}{2}K_{;\,i\,j}\sigma^i\sigma^j
+K_{,\,i\,\bar{j}}\sigma^i\bar\sigma^{\bar{j}} +\frac{1}{2}
K_{;\,\bar{i}\,\bar{j}}\bar\sigma^{\bar{i}}\bar\sigma^{\bar{j}}+
\ldots~,
\end{array}
\end{equation}
where subscript semicolon means covariant derivatives (\ref{kcov}).
The expansion of (anti) holomorphic superpotentials $P(\Phi)$,
$\bar{P}(\bar\Phi)$ can be also written down:
\begin{equation}\label{pex}
\begin{array}{c}
P(\phi+\pi)=P(\phi)+
P_{,i}(\phi)\sigma^{i}+\frac{1}{2}P_{,i;j}(\phi)\sigma^i\sigma^j +
\ldots ,\\
\bar{P}(\bar\phi+\bar\pi) =\bar{P}(\bar\phi)+
\bar{P}_{,\bar{i}}(\bar\phi)\bar{\sigma}^{\bar{i}}+\frac{1}{2}\bar{P}_{,\bar{i};\bar{j}}(\bar\phi)\bar{\sigma}^{\bar{i}}\bar{\sigma}^{\bar{j}}
+ \ldots
\end{array}
\end{equation}

It should be noted that the equation (\ref{geod}) for
(\ref{lamb_exp}) defines a transformation to holomorphic normal
coordinates. However the superfields $\sigma^i$ lose the chirality
properties $\bar{D}_{\dot\alpha}\pi^{i}=
\bar{D}_{\dot\alpha}\sigma^i-\Gamma^i_{jk}\bar{D}_{\dot\alpha}\sigma^j\sigma^k
+\frac{1}{2}R^i_{jk\bar{l}}\bar{D}_{\dot\alpha}\bar\phi^{\bar{l}}\sigma^j\sigma^k
+... =0$. That means, the deviation from chirality is quadratically
in quantum superfields $\sigma$. It leads to higher powers of
$\sigma$ in expansion of action and hence gives contribution beyond
one loop. Therefore for one-loop calculations we can consider the
$\sigma$ as chiral superfiled.

However, that presented above background-quantum splitting is not a
single one because except of widely used scalar multiplet
representation by means of chiral scalar superfield there are the
representations by means of chiral spinor superfield and by means of
unconstrained complex scalar superfield prepotential. All mentioned
representations are classically equivalent. Their quantum
equivalence was studied in \cite{kvek} (see also \cite{bk}).

\subsection{Unconstrained field expansion}
We already pointed out that manifestly supersymmetric expansion of
the action (\ref{kahl_act}), preserving the chirality properties
on the base of Riemann or K\"ahler normal coordinates,  is
impossible in principle. Now we provide expansion of the K\"ahler
potential using unconstrained complex scalar superfield
prepotential $U^i(\lambda)$ (see Ref. \cite{16}) and compare the
result with described above normal coordinate expansion. Let the
prepotential $U^i(\lambda)$ is such that $\Phi^i=\bar{D}^2 U^i$
and defined by the equation of parallel transport for
$\frac{dU^i}{d\lambda}$ along arbitrary non-geodesic curves
$\Phi^i(\lambda)$:
\begin{equation}
\frac{d^2U^i}{d\lambda^2}+
\Gamma^i_{jk}(\Phi,\bar\Phi)\frac{d\Phi^j}{d\lambda}\frac{dU^k}{d\lambda}
=0~.
\end{equation}
This equation can be solved subject to the initial condition
$U^i(\lambda=0)=U^i_{backgr}$,
$\frac{dU^i}{d\lambda}(\lambda=0)=\xi^i$ where $U^i_{backgr}$ is a
background prepotential and $\xi^i$ is a quantum field which again
is an unconstrained superfield. Explicitly:
\begin{equation}\label{u_ex}
U^i=U^i_{backgr}+\lambda \xi^i
-\frac{\lambda^2}{2}\Gamma^i_{jk}\bar{D}^2 \xi^j
\xi^k+...\end{equation} We observe that all higher order terms in
the expansion (\ref{u_ex}) involve the background field via $\phi$
(not $U_{backgr}$) so that the substitution of (\ref{u_ex}) into the
K\"ahler potential will yield a Lagrangian which depends on $\phi^i$
and the quantum field $\xi^i$. Though the chiral normal coordinates
do not exist, it is easy to show that this expansion is
reparametrization covariant. The expansion coefficients at all
orders in the quantum field $\xi^i$ are constructed from geometrical
objects, which are functions of the background field $\phi^i$. The
leading terms in the expansion $\Phi^i$ are given by
$$
\Phi^i(\lambda)=\phi^i +\pi^i=\phi^i +\lambda\bar{D}^2 \xi^i
-\frac{\lambda^2}{2}\bar{D^2}(\Gamma^i_{jk}\bar{D}^2\xi^j \xi^k)
+...
$$
and for K\"ahler potential we obtain
\begin{equation}\label{Kzeta}
K(\phi+\pi,\bar\phi+\bar\pi)=K +K_{,i}\bar{D}^2\xi^i
-K_{,i}\frac{1}{2}\bar{D}^2
(\Gamma^i_{jk}\bar{D}^2\xi^j\xi^k)+K_{,\bar{i}}{D}^2\bar\xi^{\bar{i}}
-K_{,\bar{i}}\frac{1}{2}{D}^2
(\Gamma^{\bar{i}}_{\bar{j}\bar{k}}{D}^2\bar\xi^{\bar{j}}\bar\xi^{\bar{k}})
\end{equation}
$$+\frac{1}{2}K_{,i;j}\bar{D}^2\xi^i\bar{D}^2\xi^j+\frac{1}{2}K_{,\bar{i};\bar{j}}{D}^2\bar\xi^{\bar{i}}{D}^2\bar\xi^{\bar{j}}
+K_{,i;\bar{j}}\bar{D}^2\xi^i D^2\bar\xi^{\bar{j}}~.
$$
One can see
that the two expansions (\ref{kex}) and (\ref{Kzeta}) are
coordinated at the one-loop level and therefore it is no need to
worry about non-chirality of the quantum superfields $\sigma^i$.
Further we will use the chiral superfields approach \cite{bk},
\cite{14}, \cite{sk}.

To conclude this subsection we point out that though the above
procedure is appropriate to construct a covariant background field
expansion, use of the unconstrained prepotential means that the
action (\ref{Kzeta}) is invariant under quantum field gauge
transformation $\delta(\xi^i -\frac{1}{2}\Gamma^i_{jk}\bar{D}^2
\xi^j \xi^k+...)=\bar{D}^{\dot\alpha}\omega^i_{\dot\alpha} , \quad
\delta \phi^i=0$. That means we have to impose the gauges and
introduce the corresponding ghosts following a quantization scheme
for gauge theories with linearly dependent generators \cite{bv}. The
treatment of an infinite tower of ghosts for a nonlinear sigma-model
defined in terms of the nonminimal scalar multiplet has been carried
out in \cite{zan} and it was found that the classical duality of the
formulations in terms of chiral scalar superfields and in terms of
complex general scalar superfileds takes place at least at the
one-loop level.

\subsection{The algebra of the covariant derivatives}
In this section we consider algebra of covariant derivatives
(\ref{cccov}) related to general coordinate transformations
(\ref{repar}).  Covariant derivatives ${\cal D}_A$ transform every
tensor superfield again into tensor superfield. As in global
supersymmetry, spinor covariant derivatives ${\cal D}_\alpha$ and
$\bar{\cal D}_{\dot\alpha}$ generate the full superalgebra of
covariant derivatives ${\cal D}_A$. We demonstrate that this
algebra is equivalent to the algebra of covariant derivatives for
 SYM theory. This fact allows us to use the general methods
developed for finding the one-loop effective action in SYM theory.

Components of curvature tensor in K\"ahlear geometry satisfies the
relation $R^i_{\,jkl} \equiv 0$. It leads to the following
property for anticommutators of covariant derivatives
\begin{equation}
\{{\cal D}^\alpha , {\cal D}^\beta\} = \{\bar{\cal D}^{\dot\alpha}
, \bar{\cal D}^{\dot\beta}\} =0~.
\end{equation}
This relation can be treated as a representation-preserving
constraint that make possible the existence of chiral scalar
superfield $\bar{\cal D}_{\dot\alpha}A=0$. The other
anticommu\-tation relations as conventional constraints means a
definition for the vector component of the superconnection
\begin{equation}
\begin{array}{l}
\{{\cal D}^\alpha, \bar{\cal D}^{\dot\alpha}\}A^i =i{\cal
D}^{\alpha\dot\alpha i}_{\;\;\;\; k}A^k~, \quad {\cal
D}^{\alpha\dot\alpha i}_{\;\;\;\; k}=
\partial^{\alpha\dot\alpha}\delta^i_k -i\bar{\cal
D}^{\dot\alpha}(\Gamma^i_{jk}D^\alpha \phi^j) ~,\\[2mm]
\{{\cal D}^\alpha, \bar{\cal D}^{\dot\alpha}\}A^{\bar{i}} =i{\cal
D}^{\alpha\dot\alpha \bar{i}}_{\;\;\;\; \bar{k}}A^{\bar{k}}~,\quad
{\cal D}^{\alpha\dot\alpha \bar{i}}_{\;\;\;\;
\bar{k}}=\partial^{\alpha\dot\alpha}\delta^{\bar{i}} _{\bar{k}}
-i{\cal D}^\alpha
(\Gamma^{\bar{i}}_{\bar{j}\bar{k}}\bar{D}^{\dot\alpha}\bar\phi^{\bar{j}})~.
\end{array}
\end{equation}
The commutators of the covariant derivatives define the spinor
superfield strengths
\begin{equation}
\begin{array}{l}
[{\cal D}_\beta, {\cal D}^{\alpha\dot\alpha}]A^{\bar{i}}
=i\delta^\alpha_\beta {\cal D}^2
(\Gamma^{\bar{i}}_{\bar{j}\bar{k}}\bar{D}^{\dot\alpha}\bar\phi^{\bar{j}})A^{\bar{k}}=-\delta^\alpha_\beta
\bar{\cal W}^{\dot\alpha \bar{i}}_{\;\;\;
\bar{k}}A^{\bar{k}}~,\\[2mm] [\bar{\cal D}_{\dot\beta}, {\cal
D}^{\alpha\dot\alpha}]A^i =i\delta^{\dot\alpha}_{\dot\beta}
\bar{\cal D}^2
(\Gamma^i_{jk}D^\alpha\phi^j)A^{k}=-\delta^{\dot\alpha}_{\dot\beta}
{\cal W}^{\alpha i}_{\;\;\;k}A^{k}~.
\end{array}
\end{equation}
and the conjugation properties:  $$[{\cal D}_\beta, {\cal
D}^{\alpha\dot\alpha}]A^i=g^{i\bar{i}}g_{k\bar{k}}{\cal
W}^{\dot\alpha \bar{k}}_{\;\;\;\bar{i}}A^k, \quad [\bar{\cal
D}_{\dot\beta}, {\cal
D}^{\alpha\dot\alpha}]A^{\bar{i}}=g_{k\bar{k}}g^{i\bar{i}}\delta^{\dot\alpha}_{\dot\beta}{\cal
W}^{\alpha k}_{\;\;\;i}A^{\bar{k}}~.$$ Finally, a commutation
relations of vector derivatives gives the definition for the
vector component of the strength superfield:
\begin{equation}
i[{\cal D}_{\beta\dot\beta}, {\cal
D}^{\alpha\dot\alpha}]=-\frac{1}{2}\delta^{\dot\alpha}_{\dot\beta}{\cal
D}_{(\beta}{\cal W}^{\alpha)} - \frac{1}{2}\delta^{\alpha}_{\beta}
\bar{\cal D}_{(\dot\beta}\bar{\cal
W}^{\dot\alpha)}=-\frac{1}{2}G^{\alpha\dot\alpha}_{\beta\dot\beta}~.
\end{equation}
Thus all the superfield strengths for the theory are expressed in
terms of spinor superfields
\begin{equation}\label{WR}
\begin{array}{l}
{\cal W}^{\alpha i}_{\;\; k}= -i\bar{\cal D}^2
(g^{i\bar{m}}D^\alpha
g_{\bar{m}k})=R^i_{jk\bar{l}}\partial^{\alpha\dot\alpha}\phi^j
\bar{D}_{\dot\alpha}\bar\phi^{\bar{l}} +iR^i_{jk\bar{l}}D^\alpha
\phi^j \bar{\cal D}^2\bar\phi^{\bar{l}}+\frac{i}{2}\bar{\cal
D}_{\dot\alpha}R^i_{jk\bar{l}}D^\alpha \phi^j
\bar{D}^{\dot\alpha}\bar\phi^{\bar{l}}~,\\
\bar{\cal W}^{\dot\alpha \bar{i}}_{\;\;\bar{k}}= -i{\cal D
}^2(g^{\bar{i}m}
\bar{D}^{\dot\alpha}g_{m\bar{k}})=R^{\bar{i}}_{\bar{j}\bar{k}{l}}\partial^{\alpha\dot\alpha}\bar\phi^{\bar{j}}
{D}_{\alpha}\phi^{l}
+iR^{\bar{i}}_{\bar{j}\bar{k}{l}}\bar{D}^{\dot\alpha}
\bar\phi^{\bar{j}} {\cal D}^2\phi^{l}+\frac{i}{2}{\cal
D}_{\alpha}R^{\bar{i}}_{\bar{j}\bar{k}l}\bar{D}^{\dot\alpha}
\bar\phi^{\bar{j}} {D}^{\alpha}\phi^{l}~.
\end{array}
\end{equation}
The spinor superfield strengths ${\cal W}^\alpha$ and $\bar{\cal
W}^{\dot\alpha}$ evidently obey the Bianchi identities
\begin{equation}
\bar{\cal D}_{\dot\alpha}{\cal W}_\alpha=0, \quad {\cal
D}_{\alpha}\bar{\cal W}_{\dot\alpha}=0, \quad {\cal D}^\alpha{\cal
W}_\alpha +\bar{\cal D}^{\dot\alpha}\bar{\cal W}_{\dot\alpha}=0~.
\end{equation}

Using the above covariant derivatives we introduce two basic
covariant differential operators acting on covariantly
(anti)chiral superfields. These operators  are obtained by
covariantization of the identity $\bar{D}^2 D^2 \Phi=\Box \Phi$
(where $\Box$ denotes the free d'Alembertian)
\begin{equation}\label{box+}
\Box_{+}A^i=\bar{\cal D}^2{\cal D}^2A^i =\Box_{cov} A^i +i{\cal
W}^{\alpha i}_{\;\; k}{\cal D}_\alpha A^k -\frac{i}{2}({\cal
D}_\alpha {\cal W}^{\alpha i}_{\;\; k})A^k~,
\end{equation}
where $\Box_{cov}=\frac{1}{2}{\cal D}^{\alpha\dot\alpha}{\cal
D}_{\alpha\dot\alpha}$ and $A^i$ is covariantly chiral superfield.
Analogously for covariantly antichiral superfield $A^{\bar{i}}$ ones
get
\begin{equation}\label{box-}
\Box_{-}A^{\bar{i}}={\cal D}^2\bar{\cal D}^2
A^{\bar{i}}=\Box_{cov}A^{\bar{i}} +i\bar{\cal
W}^{\dot\alpha\bar{i}}_{\;\;\bar{k}}\bar{\cal
D}_{\dot\alpha}A^{\bar{k}}-\frac{i}{2}(\bar{\cal
D}_{\dot\alpha}\bar{\cal W}^{\dot\alpha
\bar{i}}_{\;\;\bar{k}})A^{\;\;\bar{k}}~.
\end{equation}
The operators $\Box_{+}$ and $\Box_{-}$ obey the useful properties
$$
{\cal D}^2\Box_{+}=\Box_{-}{\cal D}^2, \quad \bar{\cal
D}^2\Box_{-}=\Box_{+}\bar{\cal D}^2~. $$ As we will see further
the operators (chiral and antichiral d'Alambertians) $\Box_{+}$
and $\Box_{-}$ play the crucial role for calculating the effective
action.

We see that the strength superfields ${\cal W}, \bar{\cal W}$
demonstrate the properties similar to the SYM superfield strength
properties. It is useful to compare spinor strengths definition
for generic chiral superfiled model with the definition of the
superstrengths  for the supersymmetry gauge model $W_\alpha=i\bar{
D }^2({\rm e}^{-V} D_{\alpha}{\rm e}^{V})$. One can note that the
metric of the sigma model plays the same role as a prepotential
for a gauge model. Then taking into account $U(n)$ gauge
transformations of metric $g \rightarrow {\rm
e}^{\bar\Lambda(\bar\phi)} g {\rm e}^{-\Lambda(\phi)}$, we see
that a corresponding connection should be defined as
$g^{-1}D_\alpha g$. Such a connection will have the gauge
transformation ${\rm e}^\Lambda {\cal D}_\alpha {\rm
e}^{-\Lambda}.$ Therefore one can impose "Wess-Zumino" gauge and
in particular for the metrics this means
$$
g_{i\bar{j}}(\phi,\bar\phi)= g_{i\bar{j}}(0) +
R_{i\bar{j}k\bar{l}}(0) \phi^k\bar\phi^{\bar{l}} +\ldots
$$
Such a
choice of gauge fixing is nothing but the K\"ahler normal
coordinate expansion \cite{11n}.

Thus, the obtained algebra of covariant derivatives is analogous
to the algebra of covariant derivatives for ${\cal N}=1$ SYM
theory and then we can use the powerful results developed for
quantum SYM theory \cite{bk}, \cite{14}. But it should be kept in
mind that many important results for SYM one-loop effective action
were found in the constant background field approximation
(constant strength approximation). However this approximation is
not very appropriate and interesting for the model under
consideration since it effectively means that ether all background
fields  $\Phi~, \bar{\Phi}$ are constant or
$R_{i\bar{j}k\bar{l}}=0$ and, therefore, a geometrical character
of the model disappears.

\section{One-loop calculations}

We define the effective action $\Gamma[\phi,\bar\phi]$ on the base
of background field generating functional $Z[\phi,\bar\phi]={\rm
e}^{\frac{i}{\hbar}\Gamma[\phi,\bar\phi]}$ by integrating over the
quantum fluctuations $\sigma^I$, $I=\{i, \bar{i}\}$. This definition
leads at one loop to
\begin{equation}\label{one-loop_def}
\begin{array}{l}
\displaystyle Z[\phi]={\rm e}^{i\Gamma[\phi]}={\rm e}^{iS_0}\int
D\sigma \det{}^{1\over 2} \left(g_{i\bar{j}}(\phi,
\bar{\phi})\right) {\rm e}^{i\int \sigma^I {\cal
H}_{IJ}\sigma^J}\\
\displaystyle ={\rm e}^{iS_0}\mbox{Det}^{-\frac{1}{2}}[{\cal H
}_I^J]~,
\end{array}
\end{equation}
where ${\cal H}_{I}^{J}= {\cal H}_{IK}g^{KJ}$, while ${\cal H}_{IK}$
is the second functional derivatives of the action (\ref{kahl_act})
over quantum fields
\begin{equation}\label{h}
{\cal H}_{IJ}= {\delta^2
S(\Phi,\bar{\Phi})\over\delta\sigma^I\delta\sigma^J}~.
\end{equation}
In the previous sections we obtained background-quantum splitting
(\ref{kex}, \ref{pex}) for the classical action (\ref{kahl_act}).
It allows us to calculate the above functional derivatives and
find the operator ${\cal H}_{IK}$ in an explicit form.

\subsection{One-loop reparametrization invariant counterterms}
In this subsection we find the divergent part of one-loop
effective action $\Gamma^{(1)}$. This functional is expressed in
terms of functional determinant of the operator (\ref{h})
$$
{\cal H}=\pmatrix{\frac{\delta^2
S}{\delta\sigma^i(z)\delta\sigma^j(z')}&\frac{\delta^2
S}{\delta\sigma^i(z)\delta\bar\sigma^{\bar{j}}(z')}\cr\frac{\delta^2
S}{\delta\bar\sigma^{\bar{i}}(z)\delta\sigma^j(z')}&\frac{\delta^2
S}{\delta\bar\sigma^{\bar{i}}(z)\delta\bar\sigma^{\bar{j}}(z')}}=\pmatrix{{\cal
H}_{++}(z,z')&{\cal H}_{+-}(z,z')\cr{\cal H}_{-+}(z,z')&{\cal
H}_{--}(z,z')}.
$$
The two-point functions ${\cal H}_{\pm\pm}(z,z')$ are covariantly
chiral $(+)$ or covariantly antichiral $(-)$ with respect to the
corresponding superspace argument. The functional derivatives for
covariantly chiral (antichiral) superfields have following forms
$$
\frac{\delta \sigma^i(z)}{\delta\sigma^j(z')}=\delta^i_j \bar{\cal
D}^2 \delta^8(z-z')\equiv \delta_{+}(z,z'), \quad \frac{\delta
\bar\sigma^{\bar{i}}(z)}{\delta\bar\sigma^{\bar{j}}(z')}=\delta^{\bar{i}}_{\bar{j}}
{\cal D}^2 \delta^8(z-z')\equiv \delta_{-}(z,z')
$$ Using the expansion (\ref{kex},
\ref{pex}) ones obtain the explicit form for the matrix of the
second functional derivatives
\begin{equation}
{\cal H}_{I}^{J}= \pmatrix{{\cal M}_{i}^{\bar{j}}\bar{\cal
D}^2&\delta^j_i \bar{\cal D}^2{\cal D}^2\cr
\delta^{\bar{j}}_{\bar{i}} {\cal D}^2\bar{\cal D}^2 & \bar{\cal
M}_{\bar{i}}^{j}{\cal
D}^2}\pmatrix{\delta^8(z-z')&0\cr0&\delta^8(z-z')}~,
\end{equation}
where we have used the covariant derivatives defined in the previous
section and\footnote{ In further manipulations it is useful to take
into account the relations
$$
\bar{\cal D}^2 K_{,\,i\,;\,j}=R^l_{i\,j\,\bar{m}}(\bar{\cal D}^2
\bar\phi^{\bar{m}})K_{,\,l}+\frac{1}{2}\left(R^l_{i\,j\,\bar{m}\,,\,\bar{n}}K_{,\,l}
+R_{\bar{n}\,i\,j\,\bar{m}}
\right)\bar{D}^{\dot\alpha}\bar\phi^{\bar{n}}
\bar{D}_{\dot\alpha}\bar\phi^{\bar{m}}~,
$$
$$
{\cal D}^2 K_{\bar{i};\bar{j}} =R^{\bar{l}}_{\bar{i}\bar{j}
m}({\cal D}^2\phi^m) K_{,\;\bar{l}}+
\frac{1}{2}\left(R^{\bar{l}}_{\bar{i}\bar{j}m;n}K_{,\;\bar{l}}+R_{n\bar{i}\bar{j}m}
\right)D^\alpha\phi^m D_\alpha \phi^n~,
$$
here semicolon subscript stands for repametrization covariant
derivatives  which were defined above.}
\begin{equation}\label{M}
{\cal M}_{i}^{\bar{j}}=\bar{\cal D}^2 K_{i;\,m}g^{m\bar{j}}+
P_{,i;j}g^{j\bar{j}}~,\quad  \bar{\cal M}_{\bar{i}}^{j}= {\cal D
}^2K_{\bar{i};\,\bar{m}}g^{\bar{m}j} +
\bar{P}_{,\bar{i};\bar{j}}g^{j\bar{j}}~.
\end{equation}
It is easy to show that the ${\cal M}$ and $\bar{\cal M}$ obey
chirality properties $\bar{\cal D}_{\dot\alpha}{\cal M}=0$, ${\cal
D}_\alpha\bar{\cal M}=0$. The matrix ${\cal H}_{I}^{J}$ can be
rewritten as a product of two matrix
\begin{equation}\label{matrixprod}
\pmatrix{{\cal M}&\bar{\cal D}^2\cr{\cal D}^2&\bar{\cal
M}}\pmatrix{\bar{\cal D}^2\delta^8(z-z')&0\cr0&{\cal
D}^2\delta^8(z-z')}=\pmatrix{{\cal M}&\bar{\cal D}^2\cr{\cal
D}^2&\bar{\cal
M}}\pmatrix{\delta_{+}(z-z')&0\cr0&\delta_{-}(z-z')}~,
\end{equation}

In further transformations we act as follows (see e.g. \cite{bk},
\cite{sk}). Using the definition (\ref{box+}, \ref{box-}) and
(\ref{matrixprod}) we rewrite the one-loop correction in the form
\begin{equation}\label{1l}
\begin{array}{l}
-i\Gamma^{(1)}=\mbox{Tr}\ln\pmatrix{0&\bar{\cal
D}^2\cr {\cal D}^2&0 }+ \mbox{Tr}\ln
\pmatrix{1&\frac{1}{\Box_{+}}\bar{\cal D}^2\bar{\cal
M}\cr\frac{1}{\Box_{-}}{\cal D}^2{\cal M}&1}\\
=\frac{1}{2}\mbox{Tr}\ln
\pmatrix{\Box_{+}&0\cr0&\Box_{-}}+\frac{1}{2}\mbox{Tr}\ln\left(1-\pmatrix{\frac{1}{\Box_{+}}\bar{\cal
D}^2\bar{\cal M}\frac{1}{\Box_{-}}{\cal D}^2{\cal
M}&0\cr0&\frac{1}{\Box_{-}}{\cal D}^2{\cal
M}\frac{1}{\Box_{+}}\bar{\cal D}^2\bar{\cal M}}\right)~.
\end{array}
\end{equation}
Then using the chiral d'Alambertian properties and the
(anti)chirality properties of  ${\cal M}, \bar{\cal M}$ one
rewrites the the second term in a form which can be combined with
the first term in (\ref{1l}) and obtains the expression
\begin{equation}
\Gamma^{(1)}= \frac{i}{2}\mbox{Tr}\ln \pmatrix{\Box_{+}-\bar{\cal
D}^2{\cal D}^2\bar{\cal M}\frac{1}{\Box_{+}}{\cal
M}&0\cr0&\Box_{-}-{\cal D}^2\bar{\cal D}^2{\cal
M}\frac{1}{\Box_{-}}\bar{\cal M}}
\end{equation}
$$ =\frac{i}{2}\mbox{Tr}_{+}\ln\left(\Box_{+}-\bar{\cal D}^2{\cal
D}^2\bar{\cal M}\frac{1}{\Box_{+}}{\cal M}\right)+ \frac{i}{2}
\mbox{Tr}_{-}\ln\left(\Box_{-}-{\cal D}^2\bar{\cal D}^2{\cal
M}\frac{1}{\Box_{-}}\bar{\cal M}\right)~.
$$
Two independent (anti)chiral functional traces in the last
expression can be treated separately by expanding the logarithm in
the power series.

Further, we use the superspace  Schwinger-De Witt techniques and
explore the structure of the effective action superfunctional,
including the analysis of divergences and finite contributions. For
these goal we use the methods developed for SYM theory \cite{bk},
\cite{14} (for recent development see e.g. \cite{sk}, \cite{eff},
\cite{gra}) and a covariant expansion of the corresponding
propagator in powers of the superfield strengths ${\cal W}_\alpha$,
$\bar{\cal W}_{\dot\alpha}$ and their covariant derivatives.

First of all we study a structure of divergences. One can show
(analogous to SYM theory) that the divergences are given by the
following expression
\begin{equation}\label{trdiv}
\Gamma^{(1)}_{div}=\frac{i}{2}\mbox{Tr}\int d^6z \ln
\Box_{+}\delta^{(6)}_{+}(z-z')|-\frac{i}{2}\mbox{Tr}\int d^6z
\bar{\cal D}^2\bar{\cal M}{\cal M}\frac{1}{\Box_{-}^2} {\cal
D}^2\delta^{(6)}_{+}(z-z')| + c.c.
\end{equation}
Explicit evaluation of the divergences (\ref{trdiv}) is based on
expansion of the logarithm of the operator in the second power in
${\cal D}$ derivatives and integrates by parts in order to release
$\delta^4(\theta-\theta')$. Note that ${\cal W}$ should not be
differentiated in the divergent terms because of dimensional
reasons. We omit the the details of the calculations, just note
that a heat kernel representation and a dimensional regularization
scheme were used \cite{bk}. It leads to a simple and compact
expression for the divergences
\begin{equation}\label{div}
\Gamma^{(1)}_{div}=-\frac{\Gamma(\omega)}{2(4\pi)^{2-\omega}}\left(\frac{m}{\mu}\right)^{-2\omega}\left(\int
d^6z \frac{1}{2}{\cal W}^{\alpha i}_{\;\; k}{\cal W}_{\alpha
i}^{\;\; k} -\mbox{tr}\int d^8z \, \bar{\cal M}{\cal M} +
c.c.\right)~,
\end{equation}
where $m$, $\mu$ are IR and UV the mass scales respectively and
$\omega=(4-d)/2$ is a regularizaton parameter. It should be noted
that this result is valid for arbitrary K\"ahler potential and
superpotential. Such form of for the one-loop effective action
looks like a supersymmetric version of the known result \cite{bb}
of Boulware and Brown.

Let us analyze the structure of the obtained divergent
contributions. First of all we point out that the term in
(\ref{div}) which is given by integral over chiral subspace can be
written in form of the $4D$, ${\cal N}=1$ supersymmetric ungauged
WZNW action \cite{swzw}:
\begin{equation}\label{swznw}
\int d^8z\, \Gamma^i_{jk}(D^\alpha
\phi^j)\left[R^k_{li\bar{m}}\bar{D}^{\dot\alpha}\bar\phi^{\bar{m}}
i\partial_{\alpha\dot\alpha}\phi^l +R^k_{li\bar{m}} \bar{\cal
D}^2\bar\phi^{\bar{m}}D_\alpha\phi^l +\frac{1}{2}\bar{\cal
D}^{\dot\alpha}
R^k_{li\bar{m}}\bar{D}_{\dot\alpha}\bar\phi^{\bar{m}}D_\alpha\phi^l\right]
+c.c.~.
\end{equation}
Moreover, the first term in (\ref{swznw}) has a form similar to the
manifestly supersymmetric expression of WZNW term proposed by
Nemeschansky and Rohm in Ref.\cite{nr}
\begin{equation}\label{NR}
S_{WZNW}=ic\int d^8z\,
\left(\beta_{ij\bar{k}}(\phi,\bar\phi)D^\alpha \phi^i
\partial_{\alpha\dot\alpha}\phi^j
\bar{D}^{\dot\alpha}\bar\phi^{\bar{k}} +c.c.\right)~,
\end{equation}
where the bosonic parts consist of the bosonic WZNW term and an
additional four derivative term. Note that in contrast to the
earlier analysis of supersymmetric WZNW term \cite{nr} where an
infinite number of unspecified constants appeared in calculations of
matrix elements based upon the N-R WZNW action (\ref{NR}) our action
(\ref{swznw}) is completely expressed only in terms of well defined
geometric quantities.

It is known that for the higher derivative terms in this form of
supersymmetric WZNW action a serious problem appears: the
auxiliary fields became dynamical. In Ref. \cite{aux} a
possibility to eliminate derivative terms of the auxiliary fields
was examined and it was found that the condition for disappearance
of these terms is equivalent to a condition of the term (\ref{NR})
vanishing $\beta_{ij\bar{k}, \bar{l}} - \beta_{\bar{k}\bar{l}i, j}
= 0$. Another possibility to overcome this problem was considered
by Gates and his collaborators who suggested a new
non-conventional form of the supersymmetric WZNW term consisting
in doubling the chiral superfields to chiral and complex linear
superfields \cite{3}. In the recent work \cite{swzw} it was
constructed the actual $4D$, ${\cal N}=1$ superspace WZNW action
related to the non-Abelian consistent anomaly
\begin{equation}
\begin{array}{rl}
S_{WZNW}=&C_0(\frac{1}{4\pi^2}) {\cal R}e \int d^8z\, \left({\cal
T}_{ij\bar{k}}D^\alpha \phi^i \partial_{\alpha\dot\alpha}\phi^j
\bar{D}^{\dot\alpha}\bar\phi^{\bar{k}}+ {\cal
T}_{i\bar{j}\bar{k}}D^2 \phi^i
\bar{D}^{\dot\alpha}\bar\phi^{\bar{j}}\bar{D}_{\dot\alpha}\bar\phi^{\bar{k}}\right.\\
+ &\left.{\cal T}_{ij\bar{k}\bar{l}}D^\alpha\phi^i D_\alpha\phi^j
\bar{D}^{\dot\alpha}\bar\phi^{\bar{k}}\bar{D}_{\dot\alpha}\bar\phi^{\bar{l}}
\right)~.
\end{array}
\end{equation}
Comparison this action with the obtained expression (\ref{swznw})
demonstrates one to one conformity. One can conclude that on a
general K\"ahler manifold the one-loop counterterms have the form
of supersymmetric WZNW term \cite{swzw}, while on the constant
curvature superspace and on-shell ${\cal D}^2\phi^i=0$ we get an
an expression analogous to N-R term \cite{nr}. The second term in
(\ref{div}) represents  the fourth-order supersymmetric action for
the nonlinear sigma model \cite{ket} with particular definition of
the allowed tensors $(G, A, T, H)$ \cite{ket} in terms of
geometrical quantities and superpotential.

\subsection{Finite contributions} In this section we present a
method for calculation of next to leading Schwinger-De Witt
coefficients for the one-loop effective action expansion on an
arbitrary background.

Finding the superfield effective action is based on calculations
of the chiral operator functional trace like
\begin{equation}
\mbox{Tr}_{\pm}{\cal A}=\int d^6z \, {\cal A}\,
\delta_{\pm}(z,z')~,
\end{equation}
In the model under consideration one has
\begin{equation}\label{gam}
\begin{array}{l}
\quad \quad \quad \quad \quad \Gamma^{(1)} =
\frac{i}{2}\mbox{Tr}_{+}\ln \Box_{+}
+\frac{i}{2}\mbox{Tr}_{-}\ln \Box_{-}\\
+\frac{i}{2}\mbox{Tr}_{+}\ln (1-\frac{1}{\Box_{+}} \bar{\cal D}^2
\bar{\cal M}\frac{1}{\Box_{-}} {\cal D}^2 {\cal M})
+\frac{i}{2}\mbox{Tr}_{-}\ln (1-\frac{1}{\Box_{-}} {\cal D}^2 {\cal
M}\frac{1}{\Box_{+}} \bar{\cal D}^2 \bar{\cal M})~.
\end{array}
\end{equation}
The above expression contains four terms, two of them are chiral
and two other are antichiral. Therefore it is sufficient to study
only chiral terms and use conjugation to obtain others. Let us
consider the contributions going from terms with ${\rm Tr}_{+}$
for example. The first term can be rewritten via a proper time
integral
\begin{equation}
\mbox{Tr}_{+}\ln \Box_{+}=\int^\infty_0\frac{ds}{s}\,{\rm
e}^{-sm^2}\int d^6z\, {\rm
e}^{s\Box_{+}}\delta_{+}(z,z')|_{z=z'}=\int^\infty_0\frac{ds}{s}\,
{\rm e}^{-sm^2} K_{+}(s)~,
\end{equation}
where we have introduced an IR cutoff $m$. The heat kernel has an
asymptotic expansion in powers $s$ and can be expressed as a series
\begin{equation}\label{sdw_exp}
K_{+}(s)=\frac{1}{(4\pi)^2s^2} \sum^\infty_{n=2}a_n (z)s^n~.
\end{equation}
It is known that the coefficients  $a_0=a_1=0$ and the first
non-trivial coefficient $a_2$ defines the divergences. So we know
that finite contributions can be given by integral over full
superspace. In particular it means that the contributions from any
coefficient $a_n$ with $n \geq 3$ are expressed as $\bar{\cal
D}^2$ acting on field strengths and their covariant derivatives
and, therefore, they can be transformed to a gauge invariant
superfunctional on the full superspace. It is allows us to write
the following differential equation for the kernel $K_{+}$
\begin{equation}\label{kaa_eq}
\begin{array}{l}
\frac{d K_{+}(s)}{ds}=\frac{1}{(4\pi)^2}\sum^\infty_{n=3}
(n-2)s^{n-3}a_n(z)=\int d^6z \,\bar{\cal D}^2 {\cal
D}^2{\rm e}^{s\Box_{+}}\delta_{+}(z,z')|_{z=z'}\\
=K^{\alpha}_{\;\;\alpha}(s)=\int d^8z \,{\cal D}^2{\rm
e}^{s\Box_{+}}\delta_{+}(z,z')|_{z=z'}
=\frac{1}{(4\pi)^2s^2}\sum_{n=0}^\infty s^n c_n(z)~.
\end{array}
\end{equation}
We see that it is convenient to redefine the coefficients in the
series (\ref{sdw_exp}) in the form $a_n=\frac{1}{n-2}\bar{\cal
D}^2 c_{n-1}$.

There are the various methods for evaluations of superfield heat
kernels\footnote{For earlier approaches to calculations of
superfield De Witt coefficients see e.g. \cite{b}}. Here we adopt
for our aims one of such methods \cite{eff}. First of all ones
present the covariant chiral delta-function by integral
\begin{equation}
\delta_{+}(z,z')=\bar{\cal D}^2\delta^8(z,z')I(z,z') = -
\delta^4(\zeta)\delta^2(\zeta)I(z,z')=-\int d^6 \eta\, {\rm
e}^{ip_{\alpha\dot\alpha}\zeta^{\alpha\dot\alpha}+\zeta^\alpha
\pi_\alpha}I(z,z')~,
\end{equation}
where $d^6\eta \equiv \frac{d^4p}{(2\pi)^4}d^2\pi$ and $I(z,z')$
is an operator of the parallel displacement \cite{sk}. Invariant
superintervals are defined as
\begin{equation}
\zeta^{\alpha\dot\alpha}=(x-x')^{\alpha\dot\alpha}-\frac{i}{2}\theta^\alpha
\bar\theta^{\dot\alpha '} +\frac{i}{2} \theta^{\alpha
'}\bar\theta^{\dot\alpha}, \quad \zeta^\alpha
=(\theta-\theta')^\alpha, \quad \bar\zeta^{\dot\alpha} =
(\bar\theta -\bar\theta')^{\dot\alpha}~.
\end{equation}
The resulting  $K^\alpha_{\;\;\alpha}$ from (\ref{kaa_eq}) is
rewritten using integral over momenta
\begin{equation}\label{kaa}
K^\alpha_{\;\; \alpha} (s) = \int d^8z \,{\cal D}^2{\rm
e}^{s\Box_{+}}\delta_{+}(z,z')|_{z=z'} =\int d^8z \,\int d^6\eta\,
\frac{-1}{2}X^\alpha X_\alpha {\rm
e}^{s\Delta_{+}}I(z,z')|_{z=z'}~,
\end{equation}
where
$\Delta_{+}=\frac{1}{2}X^{\alpha\dot\alpha}X_{\alpha\dot\alpha}+i
{\cal W}^\alpha X_\alpha  +\frac{i}{2}({\cal D}^\alpha {\cal
W}_\alpha)$, and operators $X_A$ defined as
\begin{equation}
X_{\alpha\dot\alpha}= ip_{\alpha\dot\alpha}+ {\cal
D}_{\alpha\dot\alpha}, \quad X_\alpha= \pi_\alpha
-\frac{1}{2}p_{\alpha\dot\alpha}(\bar\theta-\bar\theta')^{\dot\alpha}
+{\cal D}_\alpha~.
\end{equation}
One can verify that an algebra of these operators has the same
form as the algebra of covariant derivatives given in subsection
3.3:
\begin{equation}
\{X_\alpha, X_\beta\}=0~,\quad [X_\beta,
X^{\alpha\dot\alpha}]=\delta^\alpha_\beta \bar{\cal
W}^{\dot\alpha}~, \quad [X_{\beta\dot\beta},
X^{\alpha\dot\alpha}]=\frac{i}{2}G^{\alpha\dot\alpha}_{\beta\dot\beta}~,
\end{equation}
and $X A=({\cal D}A)+(-1)^{|a||X|}A X$. Note that the shift
$-\frac{1}{2}p_{\alpha\dot\alpha}(\bar\theta-\bar\theta')^{\dot\alpha}$
in $X_\alpha$ always vanishes in the coincidence limit. Since no any
$\bar{\cal D}_{\dot\alpha}$ operators appear during calculations, we
can consider all expressions in the coincidence limit from the very
beginning.

Next, expanding the exponent in (\ref{kaa}) and using the properties
of the integral over bosonic and fermionic momenta as well as the
action of the covariant derivatives on the parallel displacement
operator in the coincidence limit \cite{sk}
\begin{equation}
I(z,z')|_{\zeta=0}=1~, \; {\cal D}_\alpha I(z,z')|_{\zeta=0}=0~,
\; {\cal D}_\beta {\cal D}_\alpha I(z,z')|_{\zeta=0}=0~, \; {\cal
D}_{\alpha\dot\alpha} I(z,z')|_{\zeta=0} =0~,
\end{equation}
we obtain (after omitting contributions equal to total derivative)
for the coefficient $a_3$ the following expression
\begin{equation}\label{a_3_1}
a_3^{(1)}=-\frac{1}{8 m^2}\int d^8z \left(\frac{1}{6}({\cal
D}_\alpha {\cal W}^\beta)({\cal D}_\beta {\cal W}^\alpha)
+\frac{1}{6}(\bar{\cal D}_{\dot\alpha} \bar{\cal
W}^{\dot\beta})(\bar{\cal D}_{\dot\beta} \bar{\cal
W}^{\dot\alpha})+({\cal D}{\cal W})({\cal D}{\cal W})\right)~.
\end{equation}
However this is only one part of the result. Other finite
contributions having the same order on power $s$ go from the
second term in trace ${\rm Tr}_{+}$ of logarithm expansion up to
second order in (\ref{gam}). The results look like
\begin{equation}\label{a_3_2}
a_3^{(2)} = \frac{1}{4m^2}\int d^8z\, {\cal
D}^{\alpha\dot\alpha}\bar{\cal M}{\cal D}_{\alpha\dot\alpha}{\cal
M}~,
\end{equation}
and
\begin{equation}\label{a_3_3}
a_3^{(3)} =\frac{1}{4m^2}\int d^8z \, \bar{\cal M}{\cal
M}\bar{\cal M}{\cal M}~.
\end{equation}
The final result is a sum $a_3^{(1)}, a_3^{(1)}, a_3^{(3)}$
obtained by substitution of (\ref{WR}), (\ref{M}) into
(\ref{a_3_1})-(\ref{a_3_3}). For the partial case ${\cal D}_A
{\cal M}=0$ and ${\cal D}_{\alpha\dot\alpha}{\cal W}=0$ we obtain
known in SYM theory $G^2$ term $\Gamma^{(1)}\sim \int d^8z
\frac{1}{{\cal M}\bar{\cal M}}
G^{\alpha\dot\alpha}_{\beta\dot\beta}G_{\alpha\dot\alpha}^{\beta\dot\beta}$
\cite{gra}.

To conclude this section we point out that the theory under
consideration can be treated as a fenomenological ${\cal N}=1$
supersymmetric model following e.g. from some fundamental
superstring theory \cite{GSW} for description of low-energy
effects. In such a case the forms of K\"ahler and (anti)chiral
potentials are dictated by the fundamental theory. In particular,
finite terms in the effective action stipulated by
$a_{3}$-coefficient allow to find contributions to $S$-matrix of
six order in momenta which are determined by the forms of K\'ahler
and (anti)chiral potentials.

\section{Summary}

In this paper we developed an approach for studying the quantum
aspects of the $4D$ generic chiral superfield model. The model is
given in terms of the K\"ahler potential and chiral and antichiral
potentials. Effective action for the model under consideration is
formulated on the base of background-quantum splitting and in
one-loop approximation preserves all symmetries of the classical
theory.

We introduced the reparametrization covariant derivatives acting
on superfields and constructed their algebra in terms of
commutators and anticommutators. It was proved that structure of
this algebra coincides with ones for the covariant derivatives in
SYM theory and the K\"ahler metric plays the role analogous to the
prepotential in the SYM theory. We also constructed the chiral and
antichiral d'Alambertians. These results open the possibilities to
apply the methods, developed for evaluation of the effective
action in SYM theory, for study of the effective action in $4D$
generic chiral superfield model.

We formulated the superfield proper-time techniques for covariant
computations of the one-loop effective action. Both divergent and
leading finite  contributions to the one-loop effective action
were found in an explicit form in geometric terms. It was showed
that the divergent term reproduces the supersymmetric WZNW term
\cite{swzw} and fourth-order supersymmetric nonlinear sigma model
\cite{ket}.

\section*{Acknowledgment}
The work was supported in part by RFBR grant, project No
06-02-16346, grant for LRSS, project No 4489.2006.2 and INTAS grant.
I.L.B is grateful to DFG grant, project No 436 RUS 113/669/0-3 for
partial support. The work of N.G.P was supported in part by RFBR
grant, project No 05-02-16211.


\begin{thebibliography}{000}

\bibitem{bel}A. A. Belavin,
A. M. Polyakov and A. B. Zamolodchikov, Nucl. Phys. B  241 (1984)
333.

\bibitem{1}
L. Alvarez-Gaume, D.Z. Freedman, Phys. Lett. B 94 (1980) 171;
Commun. Math. Phys. 80 (1981) 443; L. Alvarez-Gaume, D.Z. Freedman
and S. Mukhi, Ann. of Phys. 134 (1981) 85.

\bibitem{5}L.
Alvarez-Gaume, D.Z. Freedman, Phys. Rev. D22 (1980) 846; J.
Bagger, E. Witten, Phys. Lett. B118 (1982) 103; L. Alvarez-Gaume,
P. Ginsparg, Commun.Math.Phys.102 (1985) 311; L. Alvarez-Gaume,
S.R. Coleman and P.H. Ginsparg, Commun.Math.Phys.103 (1986) 423.

\bibitem{7}B. Spence, Nucl.Phys. B260 (1985) 630; I.L. Buchbinder, S.V. Ketov,
Theor. Math. Phys. 77 (1988) 1032; Fortsch.Phys. 39 (1991) 1.

\bibitem{ket}A.A.
Deriglazov, S.V. Ketov, Theor.Math.Phys. 77 (1988) 1160; S.V. Ketov,
Quantum Nonlinear Sigma Models: From Quantum Field Theory to
Supersymmetry, Conformal Field Theory, Black Holes and Strings,
Berlin, Germany: Springer (2000).

\bibitem{bk}I.L. Buchbinder and S.M. Kuzenko, Ideas and Methods of
Supersymmetry and Supergravity or a Walk Through Superspace, IOP
Publ. Bristol and Philadelphia, 1998.


\bibitem{14}S.J. Gates, Jr., M.T. Grisaru, M. Ro\v{c}ek and W. Siegel,
Superspace, Benjamin Cummings, Reading, MA, 1983.


\bibitem{N2}A. S. Galperin, E. A. Ivanov, V. I. Ogievetsky and E. S.
Sokatchev, Harmonic Superspace, Cambridge University Press.
Cambridge, 2001.

\bibitem{z}B. Zumino, Phys. Lett. B87 (1979) 203.

\bibitem{GSW}M.B. Green, J.H. Schwarz, E. Witten, "Superstring
theory", Cambridge, UK: University Press, 1987.

\bibitem{bcp} I.L. Buchbider, M. Cveti\'c, A.Yu. Petrov,
Mod.Phys.Lett. A15 (2000) 783, hep-th/9906141; Nucl.Phys. B571
(2000) 358, hep-th/9906141; I.L. Buchbinder, A.Yu. Petrov,
Phys.Lett. B461 (2000) 209, hep-th/9905062; A. Brignole, Nucl.Phys.
B579 (2000) 101, hep-th/0001121; S.G. Nibberlink, T.S. Nyawelo, JHEP
0601 (2006) 034, hep-th/0511004.

\bibitem{wein}S. Weinberg, Physica A  96 (1979) 327; "The
quantum theory of fields", vol. II, Cambridge, UK: University Press,
2001; J. Gasser and H. Leutwyler, Ann. of Phys, 158 (1984) 142;
Nucl. Phys. B 250 (1985) 4654.

\bibitem{Mary} Mary K. Gaillard, Effective nonrenormalizable models
at one loop, preprint LBL-24114 and UCB-PTH-87/44, 1987.

\bibitem{an}J. Wess and B. Zumino, Phys. Lett. B 37 (1971) 95, E.
Witten, Nucl. Phys B 223 (1985) 422.

\bibitem{nr}D. Nemeschansky and R. Rohm, Nucl. Phys. B 249 (1985)
157; T.E. Clark and S.T. Love, Phys. Lett. B138 (1984) 289.

\bibitem{swzw} S.J. Gates, Jr., M.T. Grisaru, M.E. Knutt and S. Penati, Phys.Lett. B503 (2001) 349,
hep-ph/0012301; S. J. Gates, Jr., M. T. Grisaru, M. E. Knutt, S.
Penati and H. Suzuki, Nucl. Phys. B596 (2001) 315 hep- th/0009192;
S. J. Gates, Jr., M. T. Grisaru and S. Penati, Phys. Lett. B481
(2000) 397, hep-th/0002045.



\bibitem{3}S.J. Gates, Phys.Lett. B365 (1996) 132,
hep-th/9508153; Nucl.Phys. B485 (1997) 145, hep-th/9606109;
 S.J. Gates and S.M. Kuzenko, Nucl.Phys. B543 (1999) 122,
hep-th/9810137; Fortsch.Phys. 48 (2000) 115, hep-th/9903013.

\bibitem{kuz} S.M. Kuzenko, Int.J.Mod.Phys. A 14 (1999) 1737,
hep-th/9806147.

\bibitem{cnm}S. J. Gates, Jr, S. Penati and G.
Tartaglino-Mazzucchelli, 6D Supersymmetry, Projective Superspace and
4D, N=1 Superfields, hep-th/0508187; S. M. Kuzenko and W. D. Linch
III, JHEP 0602 (2006) 038, hep-th/0507176; G.
Tartaglino-Mazzucchelli, Phys. Lett. B599 (2004) 324,
hep-th/0404222.

\bibitem{6} J. Honerkamp, Nucl.Phys. B36 (1972) 130.


\bibitem{how}P.S. Howe and K.S. Stelle, Int.J.Mod.Phys.A4 (1989)
1871.


\bibitem{8}K. Stelle, Phys Rev. D16 (1977) 953; E. Fradkin, A. Tseytlin, Phys.
Repts. 119 (1985) 233.


\bibitem{9D2}B. Chandrasekhar, Phys. Rev. D70 (2004) 125003, hep-th/0408184;
Phys.Lett. B614 (2005) 207 , hep-th/0503116. L. Alvarez-Gaume, M.A.
Vazquez-Mozo, JHEP 0504 (2005) 007, hep-th/0503016; K. Araki, T.
Inami, H. Nakajima and Y. Saito, JHEP 0601 (2006) 109,
hep-th/0508061.


\bibitem{9}O.D. Azorkina, A.T. Banin, I.L. Buchbinder,
N.G. Pletnev, Mod. Phys. Lett. A20 (2005) 1423, hep-th/0502008;
T.A. Ryttov, F. Sannino, Phys.Rev. D71 (2005) 125004,
hep-th/0504104; S. V. Ketov, Non-anti-commutative deformation of
complex geometry, hep-th/0602066.

\bibitem{9r}I. Jack, D.R.T. Jones, L.A. Worthy, Phys.Rev. D72 (2005)
065002, hep-th/0505248; O.D. Azorkina, A.T. Banin, I.L.
Buchbinder, N.G. Pletnev, Phys. Lett. B 633 (2006) 389,
hep-th/0509193; Phys. Lett. B 635 (2006) 50, hep-th/0601045; M. T.
Grisaru, S. Penati, A. Romagnoni, JHEP 0602 (2006) 043,
hep-th/0510175.


\bibitem{11}T.E. Clark, S.T. Love, Nucl.Phys. B301 (1988) 439.

\bibitem{11n}K. Higashijima, M. Nitta, Prog.Theor.Phys.105 (2001) 243,
hep-th/0006027, Prog.Theor.Phys.108 (2002) 185, hep-th/0203081.


\bibitem{kvek}I.L. Buchbinder and S.M. Kuzenko, Nucl.Phys. B308 (1988)
162.

\bibitem{16}P.S.
Howe, G. Papadopoulos and K.S. Stelle, Phys.Lett.B174 (1986) 405.


\bibitem{sk} S.~M.~Kuzenko and I.~N.~McArthur, JHEP 0305 (2003)
015, hep-th/0302205; S.M. Kuzenko, Co\-vari\-ant super\-graphs
I-III, www.thphys.\-uni-heidel\-berg.de/~hebe\-cker/kuzenko2.pdf;
kuzenko3.pdf;


\bibitem{bv}I.A. Batalin, G.A.
Vilkovisky, Phys.Lett. B120 (1983) 166, Phys. Rev. D28 (1983) 2567.

\bibitem{zan}S. Penati, A. Refolli, A. Van Proeyen
and D. Zanon, Nucl.Phys.B514 (1998) 460, hep-th/9710166.

\bibitem{b} I.L. Buchbinder, Yad.Fiz.(Soviet Journal of Nuclear
Physics) 36, No 8 (1982) 509.

\bibitem{eff}I.N. McArthur, T.D. Gargett, Nucl.Phys. B497 (1997) 525,
hep-th/9705200; N.G. Pletnev, A.T. Banin, Phys.Rev. D60 (1999)
105017, hep-th/9811031.



\bibitem{gra}D.T. Grasso, JHEP 0211 (2002)
012, hep-th/0210146; JHEP 0409 (2004) 054, hep-th/0407264.

\bibitem{bb}D. Boulware and L.Brown, Ann. Phys. 138 (1982) 392.

\bibitem{aux}M. Nitta, Mod.Phys.Lett.A15 (2000) 2327, hep-th/0101166.



\end{thebibliography}
\end{document}